\documentclass[11pt,a4paper]{article}
\pdfoutput=1
\usepackage{jheppub}
\usepackage[hyperref,svgnames]{xcolor}
\usepackage[latin1]{inputenc}
\usepackage{amsmath}
\usepackage{amsfonts}
\usepackage{amssymb}
\usepackage{booktabs}
\usepackage{graphicx}
\usepackage{setspace}
\usepackage{mathrsfs}
\usepackage{tikz}
\usepackage{caption}
\usepackage{subcaption}

\definecolor{hgreen}{rgb}{0,.3,0}
\definecolor{hred}{rgb}{.3,0,0}
\definecolor{hblue}{rgb}{0,0,.3}
\definecolor{LightGray}{gray}{0.95}

\numberwithin{equation}{section}

\title{Non-Canonical Q-balls\\
}

\author{Olivier Lennon}
\emailAdd{olivier.lennon@physics.ox.ac.uk}

\affiliation{Rudolf Peierls Centre for Theoretical Physics, University of Oxford, Clarendon Laboratory, Parks Road, Oxford OX1 3PU, United Kingdom}

\abstract{
Theories possessing non-canonical kinetic terms are often studied for Q-ball states in an ad-hoc manner. This paper seeks to generalise their study for both thin- and thick-wall Q-balls. Specifically, we show that theories whose potential cannot house Q-balls can do so by virtue of their non-canonical kinetic terms. Furthermore, we also constrain the theories that possess an energetically stable thick-wall limit, with ramifications for their early universe phenomenology.
}

\graphicspath{{./figs/}}

\preprint{OUTP-21-29P}

\begin{document}

\tikzstyle{every picture}+=[remember picture]
\usetikzlibrary{shapes.geometric}
\usetikzlibrary{calc}
\usetikzlibrary{decorations.pathreplacing}
\usetikzlibrary{decorations.markings}
\usetikzlibrary{decorations.text}
\usetikzlibrary{patterns}
\usetikzlibrary{backgrounds}
\usetikzlibrary{positioning}
\tikzstyle arrowstyle=[scale=2]
\tikzstyle directed=[postaction={decorate,decoration={markings,
		mark=at position 0.6 with {\arrow[arrowstyle]{>}}}}]
\tikzstyle rarrow=[postaction={decorate,decoration={markings,
		mark=at position 0.999 with {\arrow[arrowstyle]{>}}}}]

\everymath{\displaystyle}

\maketitle

\section{Introduction}

Q-balls~\cite{Coleman:1985ki} are objects that can exist within scalar field theories that are invariant under a group of transformations and are a highly studied example of a non-topological soliton~\cite{Lee:1991ax}. For a complex scalar field, whose quanta possess unit charge, a Q-ball represents a semi-classical, coherent state of $Q$ quanta -- in some sense, these can be thought of as a bound state of these particles. Originally conceived in theories of a single field possessing a $U(1)$ symmetry, the classic Q-ball analysis has been expanded to accommodate multiple fields~\cite{Kusenko:1997zq}, more complicated symmetry groups~\cite{Safian:1987pr}, as well as gauging the stabilising symmetry~\cite{Lee:1988ag, Heeck:2021zvk}.

These spherically-symmetric objects are the state within the theory that minimises the energy for a given charge, i.e., they are the most energy efficient way for a scalar theory to store charge. However, finding the extremum of the energy is not analytically tractable in the most general of scenarios, but analytic limits do exist. Specifically, these are referred to as the thin- and thick-wall limits~\cite{Coleman:1985ki, Kusenko:1997ad}, valid as descriptions for large and small charge, respectively. More recently, work has been done in Ref.~\cite{Heeck:2020bau} to analytically extend the thin-wall limit to smaller charges. In this paper, we will see that not every theory that possesses a thin-wall limit has a stable thick-wall limit, and so this work is particularly relevant.

Aside from their value as interesting theoretical objects, Q-balls have been studied phenomenologically, owing mostly to the fact that they are absolutely stable in the absence of couplings to lighter fields. As such, they have long been considered as candidates for dark matter~\cite{Kusenko:1997si, Kusenko:2001vu, Graham:2015apa, Ponton:2019hux}, as well as studied in the context of supersymmetric theories~\cite{Kusenko:1997si, Kusenko:1997zq}, or in theories of extra dimensions~\cite{Demir:2000gj, Abel:2015tca}. Thus far, Q-balls have not been discovered experimentally. However, it is believed that their signatures will be striking~\cite{Gelmini:2002ez, Kusenko:1997vp, Croon:2019rqu, White:2021hwi}.

In this work, we add to the pantheon of Q-ball literature by studying theories with non-canonical kinetic terms for stable Q-ball states. Specific theories of this type have been studied in an ad-hoc manner~\cite{Distler:1986ta, Bishara:2017otb, Bishara:2021fag}, but we seek to speak in more general terms. Consequently, we will perform the thin- and thick-wall analyses as for the ``canonical'' case -- for theories of a single, complex scalar field invariant under the group of $U(1)$ transformations -- but with terms that couple the field to its derivative. We will only consider terms with two derivatives only -- one could extend the analysis to even higher derivatives. The addition of these non-canonical kinetic terms fundamentally changes the functional dependence of the resulting expressions on the parameters of the theory, and so warrants study.

Specifically, in Section~\ref{sec:ReviewCanonical}, we review the analysis of the canonical theory of Q-balls, namely, theories of a single, complex scalar field with a $U(1)$ stabilising symmetry and \textit{no} terms that couple the field to its derivative. We demonstrate that the determination of the spatial profile of these objects is analogous to the solution of a bounce equation, which cannot be solved analytically in general. We then specialise to the thin- and thick-wall limits. In the latter, we constrain the theories that may possess stable Q-balls -- a result known in the literature, but we prove it here. We find that theories that possess a stable thin-wall limit do not generically possess a stable thick-wall limit, and this has ramifications phenomenologically in issues of formation of the early universe. In Section~\ref{sec:Non-CanQ-ball}, we perform the same analysis as before, but in theories that couple two derivatives to functions of the field. In particular, we show that in thin-wall limit, theories whose potential cannot house Q-balls can do so by virtue of their non-canonical kinetic terms. Moreover, we also similarly constrain the theories that possess a stable thick-wall limit. 

\section{Review of Canonical Q-balls}
\label{sec:ReviewCanonical}

In this section, we review the classic studies of Coleman~\cite{Coleman:1985ki} and Kusenko~\cite{Kusenko:1997ad}, namely, that of thin- and thick-wall Q-balls, respectively. We will consider theories of a single, complex scalar field with a single global $U(1)$ symmetry and canonical kinetic terms. We will refer to the Q-balls found in this class of theory as ``canonical''.

\subsection{The Minimisation Procedure}
\label{sec:CanonicalMinimisation}

We consider the theory of a single complex scalar field, $\Phi(\vec{x},t)$, and its complex conjugate, governed by the Lagrangian density
\begin{equation}
\mathcal{L} = \partial_{\mu} \Phi^{*} \partial^{\mu} \Phi - U(\Phi^{*}, \Phi).
\end{equation}
The function $U(\Phi^{*}, \Phi)$ is a potential that we leave generic for the time being -- we only stipulate that it is a function of the fields $\Phi^{*}$ and $\Phi$ only, and not their derivatives, and, without loss of generality, that the zero of the potential occurs with vanishing field. We further assume that the leading order term in the potential is quadratic in the fields, i.e., there are no terms linear or inverse in powers of the fields. The Euler-Lagrange equation for this theory is given by
\begin{equation}
\label{eq:CanonicalQballs:E-Leqns}
\partial_{\mu}\partial^{\mu} \Phi + \frac{\partial U}{\partial\Phi^{*}} = 0,
\end{equation}
with a similar equation governing the evolution of $\Phi^{*}$.

We require that this theory be invariant under some global $U(1)$ symmetry, made manifest by the invariance of the Lagrangian under the transformation
\begin{equation}
\label{eq:CanonicalQballs:SymmetryRequirement}
\Phi \to e^{i\alpha} \Phi, \quad \alpha\in\mathbb{R},
\end{equation}
with the transformation for the complex conjugate field being the complex conjugate of this transformation. Due to the Abelian nature of this global symmetry, we are free to normalise charges as we please -- we have implicitly set the charges of individual quanta of the field $\Phi$ to be unity, with the complex conjugate field being assigned the same charge with opposite sign. To this invariance is associated a conserved Noether current density,
\begin{equation}
\label{eq:NoetherCurrentDensity}
j^{\mu}  = i\left( \Phi \partial^{\mu} \Phi^{*} - \Phi^{*} \partial^{\mu} \Phi \right).
\end{equation}
This symmetry fixes the potential to be a function of the product of the fields only.

We wish to analyse this theory for Q-ball solutions. These are the states that minimise the energy per unit Noether charge. The Hamiltonian, coincident with the energy, for this theory is
\begin{equation}
\label{eq:CanonicalQballs:Ham}
H = \int\mathrm{d}^{3}x \left[ \dot\Phi^{*} \dot\Phi + \vec\nabla \Phi^{*} \cdot \vec\nabla \Phi + U(\Phi^{*},\Phi) \right].
\end{equation}
To determine the states with the lowest energy for a given charge, we employ the method of Ref.~\cite{Kusenko:1997zq}. We introduce a Lagrange multiplier, $\omega$, and minimise the functional given by
\begin{equation}
\mathcal{E}_{\omega} = H + \omega \left( Q - \int\mathrm{d}^3x \, j^{0} \right),
\end{equation}
where $j^{0}$ is the zero-component of the Noether current density. Note that minimisation of this functional with respect to the Lagrange multiplier yields the charge of the configuration. In our theory, this functional evaluates to
\begin{equation}
\label{eq:CanonicalQballs:Eulerian}
\mathcal{E}_{\omega} = \omega Q + \int\mathrm{d}^3x \left[ \left| \dot\Phi - i\omega\Phi \right|^2 - \omega^2\Phi^* \Phi + \vec\nabla \Phi^{*} \cdot \vec\nabla \Phi + U(\Phi^{*},\Phi) \right],
\end{equation}
where we have completed the square on the terms with explicit time-dependence. The solution $\Phi = 0$ corresponds to a configuration with $Q = 0$, and so we infer that a configuration with finite charge must have a profile that differs from zero over some finite domain.

The first term under the integral contains all the explicit time-dependence and is positive semi-definite. It is minimised when it vanishes, i.e., if we choose
\begin{equation}
\label{eq:CanonicalQballs:FunctionalForm}
\Phi(\vec{x},t) = e^{i\omega t} \phi(\vec{x}),
\end{equation}
where $\phi(\vec{x})$ is, without loss of generality, a real-valued function of the spatial coordinates. This functional form is generic to Q-ball solutions: they are said to rotate in field space with an angular velocity $\omega$. Given this prescription, the Euler-Lagrange equation governing the shape of the spatial potential, derived from Eq.~\eqref{eq:CanonicalQballs:E-Leqns}, is thus
\begin{equation}
\label{eq:BounceEquation}
\nabla^2 \phi = \frac{1}{2} \frac{\mathrm{d}}{\mathrm{d} \phi}\left(U(\phi) - \omega^2 \phi^2\right),
\end{equation}
where $U(\phi)$ is the potential derived from $U(\Phi^{*}\Phi)$ where $\Phi^{*}\Phi\to\phi^2$. Notice, the full function in brackets defines a new effective potential function -- the bounce potential -- under which $\phi$ is determined. This differential equation is the well-studied bounce equation associated to the creation of bubbles of true vacua in the early universe~\cite{Coleman:1977py, Callan:1977pt, Coleman:1977th} for the potential given in brackets on the right-hand side. This differential equation is not possible to solve in general. However, analytic expressions can be found in certain limits:
\begin{itemize}
\item $Q$ is very large, and the energy is dominated by the volume: these are the thin-wall Q-balls first appearing in Ref.~\cite{Coleman:1985ki};
\item $Q$ is small, and surface effects are an important energy contribution: these are the thick-wall Q-balls first appearing in Ref.~\cite{Kusenko:1997ad}.
\end{itemize}
We study both of these limits below. What is known, however, is that the lowest energy configuraton is always spherically symmetric~\cite{Coleman:1977th}, i.e.,
\begin{equation}
\label{eq:CanonicalQballs:SphericalProfile}
\phi(\vec{x}) = \phi(r).
\end{equation}
This differential equation is then solved with the following boundary conditions:
\begin{equation}
\label{eq:CanonicalQballs:BCs}
\lim_{r\to0}\phi = \phi_0, \quad \lim_{r\to0}\frac{\mathrm{d}\phi}{\mathrm{d}r} = 0, \quad \lim_{r\to\infty}\phi = 0, \quad \lim_{r\to\infty}\frac{\mathrm{d}\phi}{\mathrm{d}r} = 0,
\end{equation}
where $\phi_0\in\mathbb{R}$ is some constant. These boundary conditions will ensure that the configuration does indeed differ from zero over some finite range, and that the energy remains finite.


\subsection{Thin-Wall Q-balls}
\label{sec:CanonicalQballs:ThinWall}

Thin-wall Q-balls correspond to the large volume limit, whereby the energy of the Q-ball is dominated by a large, homogeneous core of ``Q-matter'',\footnote{Q-matter is so named due to the scaling of the mass, derived soon, of the Q-ball with $Q$. This is similar to nuclear matter, which scales approximately with $N$. It should be further noted that Q-matter represents a state that spontaneously breaks the global symmetry which stabilises the Q-ball, in a manner similar to superfluids. The resulting NGB modes -- phonons -- are massless in the infinite volume limit, but pick up a mass in the finite volume limit, and represent the lightest excitation modes of the Q-ball.} and a spatially thin wall that interpolates between the VEV in the core, $\phi_0$, and the vacuum of the theory, $\phi = 0$. This approximation scheme was first studied in Ref.~\cite{Coleman:1985ki}. In this limit, we approximate the spatial profile by
\begin{equation}
\phi(r) \approx \phi_0 \Theta(R - r),
\end{equation}
where $\phi_0 \in \mathbb{R}$ and $R$ is the radius of the core. This ansatz is consistent with the boundary conditions given in Eq.~\eqref{eq:CanonicalQballs:BCs}. 

To determine an expression for the energy of thin-wall Q-balls, we reconsider the functional given in Eq.~\eqref{eq:CanonicalQballs:Eulerian}, which in this limit becomes
\begin{equation}
\mathcal{E}_{\omega} \approx \omega_0 Q + V\left[ -\omega_0^2 \phi_0^2 + U(\phi_0) \right],
\end{equation}
where
\begin{equation}
V = \frac{4 \pi}{3}R^3,
\end{equation}
is the volume of the core of the Q-ball. This function must be minimised with respect to $\omega_0$, $V$ and $\phi_0$. For stability, the value at the minimum must be less than $Q m_{\phi}$, as otherwise it would be energetically favourable for the Q-ball to classically decay to $Q$ quanta of the scalar field. Minimisation with respect to $\omega_0$ yields
\begin{equation}
Q = 2\omega_0 \phi_0^2 V,
\end{equation}
which is expected -- this is precisely the charge of a configuration in the thin-wall limit, as derived from the Noether current density of Eq.~\eqref{eq:NoetherCurrentDensity}. The Lagrange multiplier was, after all, initially introduced in order to fix the charge of the Q-ball. Reinserting this, we obtain the energy of the core of the Q-ball
\begin{equation}
E = \frac{Q^2}{4 \phi_0^2 V} + U(\phi_0)V.
\end{equation}
To determine the volume of the core, we minimise this expression with respect to the volume to obtain
\begin{equation}
V^2 = \frac{Q^2}{4 \phi_0^2 U(\phi_0)}.
\end{equation}
Once more, reinserting yields the rest mass of the Q-ball
\begin{equation}
m_{Q} = Q \sqrt{\frac{U(\phi_0)}{\phi_0^2}}.
\end{equation}

Finally, the energy must be minimised with respect the to field value, subject to the constraint $\omega_0 < m_{\phi}$ for classical stability. This yields
\begin{equation}
\label{eq:CanonicalQballs:FieldCondition}
\frac{\mathrm{d} U(\phi_0)}{\mathrm{d} \phi_0} = 2 \frac{U(\phi_0)}{\phi_0}.
\end{equation}
This constrains the potentials that can admit Q-ball solutions. The most straightforward example of a classical potential that admits thin-wall Q-balls is
\begin{equation}
\label{eq:ExamplePotential}
U(\Phi^*,\Phi) = m_{\phi}^2 \Phi^* \Phi - \frac{c_p}{\Lambda^{p-4}} (\Phi^* \Phi)^{p/2} + \frac{c_q}{\Lambda^{q-4}} (\Phi^* \Phi)^{q/2},
\end{equation}
where $c_p,c_q > 0$ are constant coefficients, $\Lambda$ is some mass scale, and $q > p > 2$. Upon insertion of the Q-ball ansatz in Eq.~\eqref{eq:CanonicalQballs:FunctionalForm}, and taking the thin-wall limit, this becomes
\begin{equation}
U(\phi_0) = m_{\phi}^2 \phi_0^2 - \frac{c_p}{\Lambda^{p-4}} \phi_0^p + \frac{c_q}{\Lambda^{q-4}} \phi_0^q,
\end{equation}
 The condition given in Eq.~\eqref{eq:CanonicalQballs:FieldCondition} tells us that
\begin{equation}
\phi_0^{q-p} = \left(\frac{p-2}{q-2}\right)\frac{c_p}{c_q} \Lambda^{q-p}.
\end{equation}
The volume of the resultant Q-ball is given by
\begin{equation}
V = \frac{1}{2 \Lambda^2 m_{\phi}} \left( \frac{q - 2}{p - 2} \frac{c_q}{c_p} \right)^{2/(q-p)} \left[1 - c_p \left( \frac{p-2}{q-2} \frac{c_p}{c_q} \right)^{(p-2)/(q-p)} \left( \frac{q-p}{q-2} \right) \frac{\Lambda^2}{m_{\phi}^2} \right]^{-1/2}
\end{equation}
and the rest mass is
\begin{equation}
\label{eq:CanonicalQballs:ThinWallMass}
m_Q = Q m_{\phi} \left[1 - c_p \left( \frac{p-2}{q-2} \frac{c_p}{c_q} \right)^{(p-2)/(q-p)} \left( \frac{q-p}{q-2} \right) \frac{\Lambda^2}{m_{\phi}^2} \right]^{1/2}.
\end{equation}
This is less than $Qm_{\phi}$ for all $q,p \in \mathbb{R}$ such that $q > p > 2$, and so the resultant Q-balls are classically stable against decay into $Q$ quanta of the field $\Phi$. We thus see that theories with polynomial potentials, with the correct sign assignment, generically contain stable Q-ball states within their spectra.

Finally, we note that for special values of the parameters of the potential, the mass in this case vanishes. This leads to the special case of surface-dominated thin-wall Q-balls, as studied in Ref.~\cite{Spector:1987ag}.

\subsection{Thick-Wall Q-balls}
\label{sec:Canonical:Thick-Wall}

As discussed in Ref.~\cite{Kusenko:1997ad}, Q-balls solutions can be found for a range of values of the Lagrange multiplier, $\omega$. Specifically, the range is given by $\omega_0 \lesssim \omega < m_{\phi}$, where $\omega = \omega_0$ is the thin-wall limit. The opposite limit of $\omega \to m_\phi$ corresponds to the small-field limit and, consequently, the small charge limit for Q-balls in the theory. We now analyse this latter limit and, as such, we approximate the bounce potential by an expansion in its lowest terms,
\begin{equation}
U_{\omega}(\phi) \approx \left( m_{\phi}^2 - \omega^2 \right)\phi^2 - \frac{c_p}{\Lambda^{p-4}} \phi^p + \frac{c_q}{\Lambda^{q-4}} \phi^q,
\end{equation}
where $c_p,c_q > 0$ are constant coefficients, $\Lambda$ is some mass scale, and $q > p > 2$. The signs in this potential are fixed in order to allow for a Q-ball solution to exist -- the negative sign on the second term allows for the second minimum in the bounce potential to exist, and the positive sign on the final term lifts and stabilises the potential. We will ignore the final term in our analysis. Its role is to place an upper bound on the charge of the Q-balls derived below -- see Ref.~\cite{Kusenko:1997ad} for details. Presently, we assume that the quadratic and next-to-quadratic terms are of roughly equal importance within the Q-ball. 

Before we proceed, notice that this small-field expansion corresponds exactly to a theory with a potential defined by Eq.~\eqref{eq:ExamplePotential}. We found in the previous subsection that this potential generically allows for stable thin-wall Q-balls to exist within its spectrum of states. It is therefore interesting to know whether this potential allows stable Q-balls with a small charge to form. We consider this below.

To find stable Q-ball solutions, we once more study the functional given in Eq.~\eqref{eq:CanonicalQballs:Eulerian}, i.e.,
\begin{equation}
\mathcal{E}_{\omega} = \omega Q + \int\mathrm{d}^3x \left[ \vec{\nabla}\phi \cdot \vec{\nabla}\phi + (m_{\phi}^2 - \omega^2)\phi^2 - \frac{c_p}{\Lambda^{p-4}}\phi^p \right].
\end{equation}
In what follows, we will constrain the value of $p$ that allows for stable Q-balls of this type to exist. Recall, in the thin-wall case, $p$ was unconstrained. The result in an arbitrary number of spatial dimensions was given without proof in Ref.~\cite{Postma:2001ea}. In what follows, we prove this result in ($3+1$)-dimensions. Defining the dimensionless variables,
\begin{equation}
\varphi \equiv \left[\frac{c_p}{\Lambda^{p-4}}\right]^{1/(p-2)} \frac{\phi}{(m_{\phi}^2 - \omega^2)^{1/(p-2)}} \quad \mathrm{and} \quad \xi_i \equiv (m_{\phi}^2 - \omega^2)^{1/2}x_i,
\end{equation}
leads to
\begin{equation}
\mathcal{E}_{\omega} = \omega Q + (m_{\phi}^2 - \omega^2)^{(6 - p)/(2p - 4)}\left[\frac{\Lambda^{p-4}}{c_p}\right]^{2/(p-2)}S_{\varphi},
\end{equation}
where $S_{\varphi}$ is the dimensionless integral
\begin{equation}
S_{\varphi} = \int\mathrm{d}^3\xi \left[ \vec{\nabla}_{\xi} \varphi \cdot \vec{\nabla}_{\xi} \varphi + \varphi^2 - \varphi^p \right],
\end{equation}
whose precise value will not concern us here -- for numerical values of this integral for different values of $p$, see Ref.~\cite{Linde:1981zj}, with the general trend being that the value is positive and increases for increasing $p$. Minimisation with respect to $\omega$ yields
\begin{equation}
\epsilon = \Omega \left(1 - \Omega^2 \right)^{(10 - 3p)/(2p - 4)},
\end{equation}
where $\Omega \equiv \omega/m_{\phi}$ is a dimensionless parameter and
\begin{equation}
\epsilon \equiv  \frac{Qm_{\phi}^{(2p-8)/(p-2)}}{S_\varphi} \left[\frac{p - 2}{6 - p}\right]  \left[\frac{c_p}{\Lambda^{p-4}}\right]^{2/(p-2)}
\end{equation}
is a dimensionless number. In principle, this expression can give $\Omega$ in terms of $\epsilon$. Notice, since $\Omega < 1$, we must have that $\epsilon>0$ in order for a valid solution to exist. Thus, for $p>6$, we must have that $S_\varphi < 0$ to maintain this sign assignment. However, this would seem to contravene the general trend of Ref.~\cite{Linde:1981zj} for which $S_{\varphi} > 0$ -- it should be noted, however, that the case for $p=6$ has not been determined. We, thus, will simply ignore $p>6$ here, though it will turn out to not matter below. Recognise also that we have implicitly assumed that $p \not= 6$ -- in this case, the Q-balls would need to have zero charge and thus cannot exist.

Instead of solving this for $\Omega$ in terms of $\epsilon$, we reinsert the expression into $\mathcal{E}_{\omega}$ to give
\begin{equation}
\label{eq:ThickWallEnergy}
\mathcal{E}_{\omega} = mQ \left[\Omega + \left[\frac{p - 2}{6 - p}\right] \frac{(1 - \Omega^2)}{\Omega}\right].
\end{equation}
For stable Q-balls to exist, we require that this expression be less than $m_{\phi}Q$, or that
\begin{equation}
\Omega + \left[\frac{p - 2}{6 - p}\right] \frac{(1 - \Omega^2)}{\Omega} < 1,
\end{equation}
which can be rewritten as
\begin{equation}
\frac{2(4-p)}{6-p}\Omega^2 - \Omega + \frac{p-2}{6-p} < 0.
\end{equation}
Notice already that $p=4$ implies that $\Omega>1$, which is in contradiction with our starting assumption that $\Omega<1$. Assume now that $p\not=4$ such that our requirement is given by
\begin{equation}
(\Omega - 1)\left(\Omega - \frac{p-2}{2(4-p)}\right) < 0.
\end{equation}
We thus see that this is only satisfied if $\Omega$ lies between the two roots of the equation. However, recognise that if $p>10/3$, the second root is greater than unity, and so $\Omega$ lies outside of its assumed range, with the limiting case of $p = 10/3$ leading to $\Omega = 1$, which is also forbidden. We conclude that stable Q-balls may only be formed in potentials where $2<p<10/3$, and so the only integer $p$ that allows for thick-wall Q-balls is $p = 3$, i.e., the case of study in Ref.~\cite{Kusenko:1997ad}.

This is an interesting result as it implies that theories with a stable thin-wall limit do not generically have a stable thick-wall limit. This has phenomenological ramifications for the formation of Q-balls in the early universe. Specifically, theories that form Q-balls through the aggregation of charge -- solitosynthesis~\cite{Frieman:1989bx, Griest:1989bq, Kusenko:1997hj} -- must overcome this lack of stability in the small charge limit.

\section{Q-balls in Theories with Non-Canonical Kinetic Terms}
\label{sec:Non-CanQ-ball}

In this section, we modify the analysis of the previous section by allowing our theory to have non-canonical kinetic terms. As the Lagrange multiplier introduced in the analysis enters into our analysis through the time-derivative of the field, the resulting expressions are functionally different, and thus are worthy of study.

\subsection{The Minimisation Procedure}
\label{sec:QballsDerivative:Minimisation}

We now incorporate a coupling between the field and its derivative,
\begin{equation}
\mathcal{L} = \left(1 + f(\Phi^*, \Phi)\right) \partial_{\mu}\Phi^* \partial^{\mu}\Phi - U(\Phi^*, \Phi).
\end{equation}
The function $U(\Phi^{*},\Phi)$ is once more a potential, and $f(\Phi^{*},\Phi)$ is a function that denotes the coupling between the field and its derivative. Note, as $f$ modifies the time derivative -- the kinetic energy -- terms in the Lagrangian density, this furnishes a non-canonical kinetic term. We leave both of these functions generic for the time being, demanding only that both be functions of the fields only, and not their derivatives, and that, for simplicity, they vanish for vanishing field.\footnote{Technically, $f$ would tend to some constant as $\phi \to 0$. However, with field redefinitions, we can make sure that the coefficient of the kinetic term is still unity in this limit. We thus assume, for simplicity, that this has already been done.} The Euler-Lagrange equation for this theory is given by
\begin{equation}
(1 + f) \partial_{\mu}\partial^{\mu}\Phi + \left( \partial_{\mu}\Phi \partial^{\mu}\Phi \right) \frac{\partial f}{\partial \Phi} + \frac{\partial U}{\partial \Phi^*} = 0,
\end{equation}
with an analogous equation governing the behaviour of $\Phi^*$.

We require that this theory be invariant under some $U(1)$ symmetry, in the same way as in Eq.~\eqref{eq:CanonicalQballs:SymmetryRequirement}. We once more set the charges of the quanta of the field $\Phi$ to be unity. The Noether current density associated to $U(1)$ symmetry in this theory is then
\begin{equation}
\label{eq:QballsDerivative:Current}
j^{\mu} = i (1 + f) \left( \Phi \partial^{\mu} \Phi^{*} - \Phi^{*} \partial^{\mu} \Phi \right).
\end{equation}
This symmetry also fixes the potential and coupling function to be functions of the product of the fields.

The Hamiltonian for this theory is given by
\begin{equation}
H = \int\mathrm{d}^{3}x \left[ (1+f) \left( \dot\Phi^{*} \dot\Phi + \vec\nabla \Phi^{*} \cdot \vec\nabla \Phi \right) + U(\Phi^{*},\Phi) \right].
\end{equation}
We seek Q-ball solutions, which are the configurations of the field that minimise the energy for a fixed charge. We once more introduce a Lagrange multiplier, $\omega$, to fix the charge of the configuration, and analyse the functional
\begin{equation}
\mathcal{E}_{\omega} = \omega Q + \int\mathrm{d}^3x \left[ (1+f) \left( \left| \dot\Phi - i\omega\Phi \right|^2 - \omega^2\Phi^* \Phi + \vec\nabla \Phi^{*} \cdot \vec\nabla \Phi \right) + U(\Phi^{*},\Phi) \right].
\end{equation}
We note that, for this theory to not contain tachyons or negative probability states, the function $f$ cannot change the sign on the kinetic terms. We thus demand that $f>-1$. We see that, as in the canonical case, the term that contains explicit time-dependence is positive semi-definite and is minimised if
\begin{equation}
\Phi(\vec{x},t) = e^{i\omega t} \phi(\vec{x}),
\end{equation}
where $\phi(\vec{x})$ is some real-valued function of the spatial coordinates. Recall, in Section~\ref{sec:CanonicalMinimisation}, we showed that the Euler-Lagrange equation governing $\phi(\vec{x})$ was a bounce equation. However, in this case, the Euler-Lagrange equation in this theory is now
\begin{equation}
\label{eq:QballsDerivative:BounceEq}
(1+f) \nabla^2 \phi + \frac{1}{2} \frac{\mathrm{d} f}{\mathrm{d} \phi} ( \vec{\nabla} \phi \cdot \vec{\nabla} \phi ) = \frac{1}{2} \frac{\mathrm{d}}{\mathrm{d}\phi} \left(U - \omega^2 (1+f) \phi^2 \right).
\end{equation}
This equation is not in the form of a bounce equation, and so we cannot state that, in general, the lowest energy configuration of the field is spherically symmetric. However, consider the offending terms,
\begin{equation}
f\,\nabla^2\phi \quad \mathrm{and} \quad \frac{1}{2} \frac{\mathrm{d} f}{\mathrm{d} \phi} ( \vec{\nabla} \phi \cdot \vec{\nabla} \phi ).
\end{equation}
We note that the function $f(\phi)$ must be dimensionless, and thus must be suppressed by some mass scale. Furthermore, these terms will be subleading to all other terms in the regime where $\phi$ is slowly varying. Recall, the thin-wall limit is characterised by a core of constant field; the thick-wall limit describes small Q-balls in which the ``wall'' is a large component of their size. We thus posit that we can treat this equation as a bounce equation in these limits and solve it with the boundary conditions given in Eq.~\eqref{eq:CanonicalQballs:BCs} to give the properties of a spherically symmetric solution. Of course, in the thick-wall case in particular, it will need to be checked that this assumption holds \textit{a posteriori} in any specific theory. Since we will be speaking in general terms, we will forthwith assume that this holds -- in Ref.~\cite{Bishara:2017otb}, a theory with non-canonical kinetic terms was studied, and this assumption was found to hold in the thick-wall limit.

Under this assumption, the structure of the bounce potential is analogous to that found in Eq.~\eqref{eq:BounceEquation}, and so much of the canonical analysis of the bounce potential found in Ref.~\cite{Kusenko:1997ad} carries over. Specifically, the bounce potential in the thin- and thick-wall limits is given by
\begin{equation}
U_{\omega}(\phi) = U - \omega^2 (1+f) \phi^2.
\end{equation}
We earlier assumed, for simplicity, that $f \to 0$ when $\phi \to 0$, and so the bounce potential vanishes at zero field. Therefore, by Ref.~\cite{Kusenko:1997ad}, we still seek Q-ball solutions in the range $\omega_0 \leq \omega < m_{\phi}$, where the thin-wall case is given by $\omega = \omega_0$ and the thick-wall case corresponds to $\omega \to m_{\phi}^{-}$.

\subsection{Thin-Wall Q-balls}
\label{sec:QballsDerivative:ThinWall}

We assume once more that the Q-ball rest mass is dominated by a spherical, homogeneous core of volume $V$. The value of the field inside the core is given by $\phi_0$. We thus find that
\begin{equation}
\mathcal{E}_{\omega} \approx \omega Q + V\left[ - \left(1+f(\phi_0)\right) \omega_0^2 \phi_0^2 + U(\phi_0) \right].
\end{equation}
This function must be minimised with respect to the Lagrange multiplier, volume and field such that the resulting rest mass be less than $Q m_{\phi}$ to ensure classical stability against decay to quanta of the field. Minimisation with respect to $\omega_0$ yields
\begin{equation}
Q = 2 \left(1+f(\phi_0)\right) \omega_0 \phi_0^2 V,
\end{equation}
as expected from Eq.~\eqref{eq:QballsDerivative:Current}. Eliminating $\omega_0$ and subsequently minimising with respect to the volume yields
\begin{equation}
V^2 = \frac{Q^2}{4 \left(1+f(\phi_0)\right) \phi_0^2 U(\phi_0)}.
\end{equation}
Reinserting this expression finally gives us an expression for the rest mass,
\begin{equation}
m_{Q} = Q \sqrt{\frac{U(\phi_0)}{\left(1+f(\phi_0)\right) \phi_0^2}}.
\end{equation}
Notice that this expression is expected from the structure of the bounce potential given in Eq.~\eqref{eq:QballsDerivative:BounceEq} since, for thin-wall Q-balls, $m_Q = Q \omega_0$~\cite{Coleman:1985ki}. Furthermore, we see that $f(\phi_0)>-1$ is a necessary condition for the existence of these Q-balls. However, this was also a necessary condition for our theory t be well-behaved, and so this is automatically satisfied.

To determine the value of the field inside the Q-ball, we minimise the rest mass with respect to the field to give
\begin{equation}
\left(1+f(\phi_0)\right) \frac{\mathrm{d} U(\phi_0) }{\mathrm{d} \phi_0} = 2\frac{U(\phi_0)}{\phi_0} \left(1+f(\phi_0) \right) + U(\phi_0) \frac{\mathrm{d} f(\phi_0)}{\mathrm{d} \phi_0}.
\end{equation}
This constrains the forms of $f(\phi_0)$ and $U(\phi_0)$.

Consider the following example. Let $U(\phi_0) = m_{\phi}^2 \phi_0^2$. This is a potential that clearly does not allow Q-ball solutions in the canonical case given in Section~\ref{sec:CanonicalQballs:ThinWall}. However, in the case presented here, Q-ball solutions exist for a $\phi_0 \not= 0$ if
\begin{equation}
\frac{\mathrm{d} f(\phi_0)}{\mathrm{d} \phi_0} = 0.
\end{equation}
The rest mass of these Q-balls are then given by
\begin{equation}
m_Q = Qm_{\phi} \frac{1}{\sqrt{1+f(\phi_0)}},
\end{equation}
which is less than $Qm_{\phi}$ if $f(\phi_0)>0$. We thus see that terms coupling the field to its derivatives can allow for Q-ball solutions to exist even when the potential without these couplings does not.

\subsection{Thick-Wall Q-balls}
\label{sec:QballsDerivative:ThickWall}

As discussed in Section~\ref{sec:Canonical:Thick-Wall}, the properties of thick-wall Q-balls are determined in the small field-limit. The bounce potential in this scenario contains two separate functions of $\phi$, 
\begin{equation}
U_{\omega}(\phi) = U(\phi) - \left( 1+f(\phi) \right) \omega^2 \phi^2.
\end{equation}
Expanding each function to its lowest order term after the quadratic gives
\begin{equation}
\label{eq:QballsDerivative:BounceExpansion}
U_{\omega}(\phi) \approx \left(m_{\phi}^2 \phi^{2} - \frac{c_p}{\Lambda^{p-4}} \phi^{p} \right) - \left( 1+ \frac{c_q}{\Lambda^{q}}\phi^q \right)\omega^2\phi^{2},
\end{equation}
where $c_p,c_q\in\mathbb{R}$, $p>2$ and $q>0$ and $\Lambda$ is some mass scale. As ever, we assume that there is some higher order term that stabilises the potential at large $\phi$. There exists three regimes we can study:
\begin{itemize}
\item Case One: $p<q+2$, and so the term coming from the function $f(\phi)$ is irrelevant in the small field limit. The appropriate analysis is that found in the canonical case in Section~\ref{sec:Canonical:Thick-Wall}, provided that $c_p>0$, otherwise no thick-wall Q-balls can exist in this potential.
\item Case Two: $p>q+2$, and so the term coming from the potential $U(\phi)$ is irrelevant in the small field limit. It is plausible that Q-balls can exist provided that $c_q>0$; if $c_q<0$, then the bounce potential doesn't contain a barrier, and thus no Q-ball solution exists.
\item Case Three: $p = q+2$, and so both terms are of the same order in $\phi$. However, this is only the case if
\begin{equation}
\frac{c_{p}}{c_p} \sim \left( \frac{\omega}{\Lambda} \right)^2 \sim \left( \frac{m_{\phi}}{\Lambda} \right)^2,
\end{equation}
where we have used the fact the $\omega \to m_{\phi}^{-}$ in the thick-wall limit in the final approximation. If this is not true, then either Case One or Case Two is a more appropriate description for any resulting objects. Stable Q-balls are a possibility if
\begin{equation}
c_{p}\Lambda^2 + c_{q}\omega^2 > 0,
\end{equation}
as otherwise no barrier exists in the bounce potential -- see Ref.~\cite{Kusenko:1997ad} for a detailed discussion of the shape of bounce potentials. Since, in the thick-wall limit, $\omega \to m_{\phi}^{-}$, we can rewrite this inequality as
\begin{equation}
\label{eq:QballsDerivative:ConditionOnPotential}
c_{p}\Lambda^2 + c_{q}m_{\phi}^2 > 0,
\end{equation}
where the signs of $c_p$ and $c_q$ are not set, apart from the fact that they cannot both be negative.
\end{itemize}

We study both Case Two and Case Three below, noting that Case Three is equivalent to Case Two in the limit $c_p \to 0$ and $c_q>0$. The bounce potential is thus given by
\begin{equation}
U_{\omega}(\phi) \approx m_{\phi}^2  (1 - \Omega^2) \phi^{2} - \frac{m_{\phi}^2}{\Lambda^{p-2}} \left(\lambda + c_{q} \Omega^2 \right) \phi^{p},
\end{equation}
where, for notational convenience, we have defined $\Omega \equiv \omega / m_{\phi}$ and $\lambda \equiv c_{p}\Lambda^2/m_{\phi}^2$. The functional we wish to study for thick-wall Q-balls is then
\begin{equation}
\mathcal{E}_{\omega} = m_{\phi}Q\Omega + \int\mathrm{d}^3x \left[ \vec{\nabla}\phi \cdot \vec{\nabla}\phi + m_{\phi}^2(1 - \Omega^2)\phi^2 -  \frac{ m_{\phi}^2}{\Lambda^{p-2}} \left(\lambda + c_{q} \Omega^2 \right) \phi^{p}
\right].
\end{equation}
We may render the integral dimensionless by introducing the dimensionless variables
\begin{equation}
\varphi = \left[ \frac{\lambda + c_{q} \Omega^2}{1 - \Omega^2}\right]^{1/(p-2)} \frac{\phi}{\Lambda} \quad \mathrm{and} \quad \xi_{i} = m_{\phi} (1 - \Omega^2)^{1/2} x_i,
\end{equation}
such that
\begin{equation}
\label{eq:QballsDerivative:Case3Eulerian}
\mathcal{E}_{\omega} = m_{\phi} Q \Omega +  \frac{\Lambda^2}{m_{\phi}} \frac{(1 - \Omega^2)^{(6-p)/(2p-4)}}{\left(\lambda + c_{q} \Omega^2\right)^{2/(p-2)}} S_{\varphi},
\end{equation}
where
\begin{equation}
S_{\varphi} = \int\mathrm{d}^3\xi \left[ \vec{\nabla}_{\xi} \varphi \cdot \vec{\nabla}_{\xi} \varphi + \varphi^2 - \varphi^{p} \right]
\end{equation}
is a dimensionless integral, whose precise value is irrelevant for the current discussion -- this integral has been calculated for some values of $p$ in Ref.~\cite{Linde:1981zj}, and, for those calculated, been found to be positive and increasing in value for increasing $p$.

Minimisation of this expression with respect to $\Omega$ yields the condition
\begin{equation}
\label{eq:QballsDerivative:MinimisationCondition}
\epsilon = \Omega \frac{(1 - \Omega^2)^{(10-3p)/(2p-4)}}{\left(\lambda + c_{q} \Omega^2\right)^{p/(p-2)}} \left( c_{q} + \frac{6 - p}{4}\lambda - c_{q} \frac{p-2}{4} \Omega^2  \right),
\end{equation}
where
\begin{equation}
\epsilon \equiv \frac{p-2}{4} \frac{Q}{S_{\varphi}} \frac{m_{\phi}^2}{\Lambda^2}.
\end{equation}
Note, as $p>2$, we see that $\epsilon>0$. In Eq.~\eqref{eq:QballsDerivative:MinimisationCondition}, the term in brackets could plausibly be negative for certain values of $p$. For now, we take this term to be positive and return to this point later. We cannot solve this expression for $\Omega$ as a function of $\epsilon$ for any $p>2$. However, we can find values of $p$ that plausibly lead to energetically stable Q-balls.

Reinserting this expression into Eq.~\eqref{eq:QballsDerivative:Case3Eulerian} yields
\begin{equation}
\mathcal{E}_{\omega} = m_{\phi} Q \left[ \Omega + \frac{p-2}{4} \frac{1}{\Omega} \frac{(1 - \Omega^2) \left(\lambda + c_{q} \Omega^2\right)}{c_{q} + \dfrac{6 - p}{4}\lambda - c_{q} \dfrac{p-2}{4} \Omega^2} \right].
\end{equation}
In order for Q-balls to form that are classically stable against decay into the quanta of the scalar field, we require the term in brackets to be less than unity. This translates into the condition
\begin{equation}
(1 - \Omega) \left[ c_q\left(\frac{p-2}{2}\right) \Omega^3 +  c_q \left(\frac{p-2}{4}\right) \Omega^2 + \Omega \left( \frac{p-4}{2} \lambda - c_q \right) + \frac{p-2}{4}\lambda \right] < 0.
\end{equation}
The first term in brackets is always positive for $0<\Omega<1$. The term in square brackets is a cubic polynomial in $\Omega$ whose shape is defined by the parameters $c_q$, $\lambda$ and $p>2$.

Notice, for $\Omega = 0$, the final term is only negative if $\lambda < 0$. However, this is perfectly allowed for all $\lambda$ and $c_q$ -- the true lower bound on $\Omega$ is $\Omega_0$, as found in the thin-wall case. In order to constrain $p$, we thus only need to show that this expression is negative for $\Omega \to 1^{-}$, as this is the vicinity in which thick-wall Q-ball solutions lie. The constraint then reduces to
\begin{equation}
\left( \lambda + c_q \right) \left(\frac{3p-10}{4}\right) < 0.
\end{equation}
By the assumption given in Eq.~\eqref{eq:QballsDerivative:ConditionOnPotential}, the first term in brackets is always positive. So we are left with the requirement $2<p<10/3$, which is exactly the same as in the case for theories that do not couple the field to its derivative. Notice, for Case Two, where $\lambda\to0$ and $c_q>0$, we have the exact same condition arising.

Focussing now on the case $p = 3$, our minimisation condition given in Eq.~\eqref{eq:QballsDerivative:MinimisationCondition} becomes
\begin{equation}
\label{eq:QballsDerivative:MinimisationCondition3}
\epsilon = \Omega \frac{(1 - \Omega^2)^{1/2}}{\left(\lambda + c_{q} \Omega^2\right)^{3}} \left( c_{q} + \frac{3}{4}\lambda -  \frac{c_{q}}{4} \Omega^2  \right),
\end{equation}
where
\begin{equation}
\epsilon = \frac{1}{4} \frac{Q}{S_{\varphi}} \frac{m_{\phi}^2}{\Lambda^2}.
\end{equation}
Recall above that we required that the term in the bracket be positive in order for this condition to be valid. We may rewrite this condition as
\begin{equation}
c_q(4 - \Omega^2) + 3\lambda > 0
\end{equation}
If both $c_q>0$ and $\lambda>0$, this is certainly true. For $c_q = -|c_q| <0$ and $\lambda>0$, we find that
\begin{equation}
\Omega^2 > 4 \left( 1 - \frac{3}{4} \frac{\lambda}{|c_q|} \right).
\end{equation}
Note, if $\lambda \to 0$, this condition fails, as we would expect in this scenario. We require that $\lambda > |c_q|$ by Eq.~\eqref{eq:QballsDerivative:ConditionOnPotential} in this case. However, this is not enough to make the right-hand side negative, such that this condition would hold generically. Thus, this is a condition on these theories and sets a lower bound on $\Omega$. However, from the analysis above, we noted that $\Omega > \Omega_0$, where $\Omega_0$ is found from the thin-wall analysis, and so this condition is only relevant if
\begin{equation}
4 \left( 1 - \frac{3}{4} \frac{\lambda}{|c_q|} \right) > \Omega_0^2.
\end{equation}
For $\lambda = -|\lambda|<0$ and $c_q>0$, we find that
\begin{equation}
\Omega^2 < 4 \left( 1 - \frac{3}{4} \frac{|\lambda|}{c_q} \right).
\end{equation}
Note, if $c_q \to 0$, this condition fails, as we would expect in this scenario. We know that $\Omega < 1$. The right-hand side of this condition is greater than unity if $c_q > |\lambda|$. However, this is required by Eq.~\eqref{eq:QballsDerivative:ConditionOnPotential}, and so this condition holds generically in this case.

Though the function given in Eq.~\eqref{eq:QballsDerivative:MinimisationCondition3} is unbound in the range $0<\Omega<1$, it is bound in the range $\Omega_0 < \Omega < 1$, where $\Omega_0$ is found from considering the thin-wall limit. When $\Omega = 1$, $\epsilon = 0$ and thus $Q = 0$, i.e., no Q-ball forms, which is expected. The upper bound on $\epsilon$ is found when $\Omega = \Omega_0$, though one would expect a thick-wall description to break down before that bound is reached. Denoting the upper bound in $\epsilon$ to be $\epsilon_0$, we can conservatively state that this description is valid for
\begin{equation}
Q \ll 155.2 \left(\frac{\Lambda}{m_{\phi}}\right)^2 \epsilon_0,
\end{equation}
where we have used that $S_{\varphi}\approx 38.8$ for the case $q+2 = 3$~\cite{Linde:1981zj}. It should be noted that this upper bound is model-dependent.

\section{Summary}

In this paper, we have studied a class of theory with terms coupling the field to its derivative -- specifically those that couple the field to two derivatives of said field -- for Q-ball solutions. We have determined the physical properties of the Q-balls in the thin-wall limit. By demanding that the minimum of the energy be energetically stable against decay to the quanta of the field, we determined the subclass of theories possessing Q-balls in the thick-wall limit. It is interesting that, despite the difference in functional form with respect to $\omega$, the bounce potential still has the same requirements with regards to allowed terms that lead to stable thick-wall Q-balls. To reiterate, for a theory of a single, complex scalar field, thick-wall Q-balls can only exist if the bounce potential, in the limit of small field, has the form
\begin{equation}
U_{\omega}(\phi) = (m_{\phi}^2 - \omega^2)\phi^2 - \left(c_{p}\Lambda^2 + c_{q}\omega^2\right)\frac{\phi^p}{\Lambda^{p-2}},
\end{equation}
where $2<p<10/3$, and $c_p,c_q\in\mathbb{R}$ and $c_{p}\Lambda^2 + c_{q}\omega^2 > 0$. If both $c_p$ and $c_q$ are negative semi-definite, Q-balls cannot form in the low $Q$ limit. Note, the canonical case corresponds to the limit $c_q \to 0$ and $c_p > 0$.

We have only studied a subclass of single-field theories that possess terms coupling the field to its derivative. It would be interesting to repeat the above analysis for theories that couple the field to a higher number of derivatives. At least in the thick-wall case, it would reduce to our analysis given in this chapter if the terms with two derivatives are present in the theory, since terms with higher derivatives will likely be further suppressed by some higher mass scale.

Furthermore, since we have shown that theories that possess a thin-wall limit do not generically possess a thick-wall limit, it would be interesting to determine the lower bound on the charge for which stable Q-balls in these theories exist. We leave this to future work. This has phenomenological ramifications for production mechanisms in real-world scenarios, particularly for solitosynthesis. In particular, it would appear that the dissolution timescale would need to be slower than the aggregation timescale until enough charge has accumulated for stability to be achieved.

\section*{Acknowledgements}

OL would like to thank John March-Russell for useful comments on this work, which was completed while OL was supported by the Colleges of St John and St Catherine, Oxford. Conversations with Fady Bishara on Q-ball physics over the years have also been particularly fruitful.

\bibliographystyle{JHEP}
\bibliography{paper}
\end{document}